\documentclass[prd,twocolumn,amsfonts,amssymb,amsmath,nofootinbib]{revtex4}

\usepackage{graphicx}
\usepackage{simplewick}
\usepackage{epstopdf}
\usepackage{multirow}
\usepackage{dcolumn}
\usepackage{hyperref}
\usepackage{cases}
\usepackage{bm}
\usepackage{epsfig}
\usepackage{color}
\usepackage[normalem]{ulem}

\def\be{\begin{equation}}
\def\ee{\end{equation}}
\def\bea{\begin{eqnarray}}
\def\eea{\end{eqnarray}}

\newcommand{\bear}{\begin{eqnarray}}
\newcommand*\Bell{\ensuremath{\boldsymbol\ell}}
\newcommand{\eear}{\end{eqnarray}}

\newlength{\tskip}
\newbox\pippobox

\def\be{\begin{equation}}
\def\ee{\end{equation}}
\def\bea{\begin{eqnarray}}
\def\eea{\end{eqnarray}}

\def\9{\nabla}

\def\6{\partial}

\def\0{(0)}

\def\>{\rightarrow}

\begin{document}

\title{{\bf Can modified gravity models reconcile the tension between CMB anisotropy and lensing maps in Planck-like observations?}}

\author{Bin Hu$^1$ and Marco Raveri$^{2,3}$}
\affiliation{
\smallskip
$^{1}$ Institute Lorentz, Leiden University, PO Box 9506, Leiden 2300 RA, The Netherlands \\
\smallskip 
$^{2}$ SISSA - International School for Advanced Studies, Via Bonomea 265, 34136, Trieste, Italy \\
\smallskip
$^{3}$ INFN, Sezione di Trieste, Via Valerio 2, I-34127 Trieste, Italy}

\begin{abstract}
{\it Planck}-2015 data seem to favour a large value of the lensing amplitude parameter, $A_{\rm L}=1.22\pm0.10$, in CMB spectra. 
This result is in $2\sigma$ tension with the lensing reconstruction result, $A_{\rm L}^{\phi\phi}=0.95\pm0.04$. 
In this paper, we simulate several CMB anisotropy and CMB lensing spectra based on {\it Planck}-2015 best-fit cosmological parameter 
values and {\it Planck} blue book beam and noise specifications. We analyse several modified gravity models within the effective 
field theory framework against these simulations and find that models whose effective Newton constant is enhanced 
can modulate the CMB anisotropy spectra in a way similar to that of the $A_{\rm L}$ parameter. 
However, in order to lens the CMB anisotropies sufficiently, like in the {\it Planck}-2015 results, the growth of matter perturbations is substantially enhanced 
and gives a high $\sigma_8$ value.
This in turn proves to be problematic when combining these data to other probes, like weak lensing from CFHTLenS, that favour a smaller amplitude of matter fluctuations.
\end{abstract}

\maketitle

%
\section{Introduction}\label{Sec:Intro}
Based on the full-mission {\it Planck} observations of temperature and polarization anisotropies of the cosmic microwave background (CMB) radiation, {\it Planck}-2015 results show that the temperature and polarization power spectra are consistent with the standard spatially-flat six-parameter $\Lambda$CDM cosmology with a primordial power-law spectrum of adiabatic scalar perturbations. Hereafter we shall call this model the base-$\Lambda$CDM.  
On the other hand, the same data, especially the temperature-temperature (TT) spectrum reveals some tension with the CMB lensing deflection angle ($d$) spectrum  reconstructed from the same maps. 
In details, the lensing amplitude in CMB temperature and polarization spectra, $A_{\rm L}=1.22\pm 0.10$, is in $2\sigma$ tension with the amplitude of the CMB trispectrum reconstructed lensing deflection angle spectrum, $A^{\phi\phi}_{\rm L}=0.95\pm0.04$ while it is expected that in the base-$\Lambda$CDM model both these quantities should be equal to unity.

The {\it Planck} collaboration finds that, compared with the base-$\Lambda$CDM model, the base-$\Lambda$CDM+$A_{\rm L}$ model can reduce the logarithmic likelihood ($\Delta \chi^2=-6.1$) and provide a better fit to the data sets with \footnote{From the PLA-PR2-2015 official chains \texttt{base-Alens-plikHM-TT-lowTEB} at \url{http://pla.esac.esa.int/pla/}.} $A_{\rm L}=1.28$ or marginalized constraint $A_{\rm L}=1.22\pm 0.10$~\cite{Planck:2015xua} . More importantly, they find that there is roughly equal preference for high $A_{\rm L}$ from intermediate and high multipoles ({\it i.e.}, the \texttt{Plik} likelihood; $\Delta\chi^2=-2.6$) and from the low-$\ell$ likelihood ($\Delta\chi^2=-3.1$) with a further small change coming from the priors. 
This means that the base-$\Lambda$CDM+$A_{\rm L}$ model can provide a better fit than base-$\Lambda$CDM model against both TT and lowP data sets. 
However, the increase in $A_{\rm L}$ will induce changes on the full sets of cosmological parameters as mentioned in the reference~\cite{Planck:2015xua}. For example, compared with the base-$\Lambda$CDM fit, the scalar index, $n_s$, is increased by $1\%$, the primordial scalar spectrum amplitude, $A_s$, is reduced by $4\%$ and the effective amplitude of the TT spectrum, $A_se^{-2\tau}$, is reduced by $1\%$. 
Through the complicated relationship between parameters and their degeneracy the re-ionization optical depth parameter, $\tau$, falls to $0.060$, which is roughly in $2\sigma$ tension with {\it Planck}-2013 temperature + WMAP low-$\ell$ polarization data results of $\tau=0.089^{+0.012}_{-0.014}$.

Inspired by these observations, in this paper, we investigate whether some modifications of gravity can relieve the tension between {\it Planck} CMB anisotropy spectra and CMB lensing results. 
To do so we will simulate a tension CMB data set that resembles the tension present in the {\it Planck}-2015 results and we will try to fit the resulting power spectra with different models to have a glimpse of the changes in the parameter that arise because of this tension. 

We shall show that the modified gravity models that we consider are not capable of alleviating this kind of tension without affecting substantially the growth of structure.
These models will then have problems in fitting simultaneously other cosmological probes. 
Moreover we will show that, if the amplitude of the reconstructed CMB lensing power spectrum is in tension with the amount of CMB lensing in the anisotropy spectra, these kind of models struggle to relieve it. 
As we will elaborate later this happens because modified gravity models predict the same amplitude for the reconstructed CMB lensing potential and the lensing effect on the CMB so that the tension is usually moved to other parameters of the model.

\section{Mock data and fiducial parameters}
We analyse several modified gravity models against two sets of simulations of CMB spectra (TT,TE,EE) and 
CMB lensing spectra (dd and Td) with the fiducial cosmological parameter equal to the {\it Planck}-2015 data release best-fit values and 
{\it Planck} blue book beam and noise specifications. 
The motivation of doing these simulations are mainly two. First of all the {\it Planck}-2015 likelihood code and the corresponding spectrum data are not yet publicly available. Second, because the cosmological parameters are degenerate with each other in a complicated way, by using simulations, 
we can efficiently isolate and study the effects coming from different parameters and their combinations.

In the following section, we will briefly review the FuturCMB\footnote{\url{http://lpsc.in2p3.fr/perotto/}} package~\cite{Perotto:2006rj} that we used to produce our simulations.
The construction of the power spectrum simulations are mainly made by two essential parts, firstly the mock data generation and the likelihood construction. 

For the mock data generation we assume, for simplicity, the raw maps to be composed of Gaussian signal ($s_{\ell m}$) and a uniform Gaussian white noise ($n_{\ell m}$),
\begin{equation}
\label{cmb_signal}
a^P_{\ell m}= s^P_{\ell m}+n^P_{\ell m}\;,
\end{equation}
where the superscript $P$ stands for temperature and E-mode polarization as we do not consider B-mode polarization.
The observation phase then convolves this signal with a Gaussian beam characterized by the full width half maximum parameter $\theta_{\rm fwhm}$.
On the top of it, we also need to add a Gaussian white noise. Up to a normalization, the noise power spectrum can be approximated as 
\begin{equation}
\label{noise_spectrum}
N^{PP'}_{\ell}\equiv\langle n_{\ell m}^{P\ast}n_{\ell m}^{P'}\rangle=\delta_{PP'}\theta^2_{\rm fwhm}\sigma_{P}^2{\rm exp}\left[\ell(\ell+1)\frac{\theta_{\rm fwhm}^2}{8\log 2}\right]\;,
\end{equation}
where $\sigma_P$ models the root mean square of the instrumental noise. In this paper, we shall adopt the {\it Planck} bluebook~\cite{Planck:2006aa} beam and noise parameters. The exact values used in our simulations are listed in Table~\ref{tab:experiment}. 
The mock data generator package FuturCMB automatically computes the noise power spectra of the lensing deflection angle based on the Hu-Okamoto~\cite{Hu:2001kj} quadratic estimator algorithms from the given temperature and polarization spectra and noises.

Then, we feed the fiducial cosmological model into FuturCMB. In this paper, we generate the fiducial spectra $C_{\ell}^{TT}$, $C_{\ell}^{TE}$, $C_{\ell}^{EE}$, $C_{\ell}^{dd}$ as well as $C_{\ell}^{Td}$ from the public Boltzmann code CAMB~\cite{CAMB,Lewis:1999bs}. 
In order to mimic {\it Planck}-2015 results, the cosmological parameter used to produce the fiducial spectra are set to the best-fit values of the base-$\Lambda$CDM+$A_{\rm L}$ model to the TT+lowP and TT+lowP+lensing data set of the {\it Planck}-2015 data release (see Table~\ref{tab:fiducialcp}). With these input the FuturCMB code will generate Gaussian-distributed random fields for $a^P_{\ell m}$ and estimate the corresponding spectra $\hat C^{PP'}_{\ell}$. 
For further details about the numerical implementation of the FuturCMB code we refer the reader to~\cite{Perotto:2006rj}.

The second step is to construct the likelihood for the mock data sets. Since it is generated from a Gaussian realization, we can write the likelihood function of the data given the theoretical template as~\cite{Tegmark:1996bz}
\begin{equation}
\mathcal L(\bf a|\Theta)\propto {\rm exp}\left(-\frac{1}{2}\bf a\dagger[\bar C(\Theta)^{-1}]\bf a\right)\;,
\end{equation}
where ${\bf a}=\{a_{\ell m}^T,a^E_{\ell m},a^d_{\ell m}\}$ is the data vector, $\Theta=(\theta_1,\theta_2,\cdots)$ is the parameter vector, and $\bar C(\Theta)$ is the theoretical data covariance matrix. 
For the detailed construction of the spectrum likelihood we refer to~\cite{Lewis:2005tp,Perotto:2006rj}.

\begin{table}
\caption{\label{tab:experiment}{\it Planck} blue book instrumental specifications}
\begin{ruledtabular}
\begin{tabular}{cccccccc}
Experiment & Frequency & $\theta_{{\rm beam}}$ & $\sigma_T$ & $\sigma_P$ \\
\hline
{\it Planck}:		&   217 &   5.02  & 13.1 & 26.7 \\
			&   143 &   7.30  & 6.0  & 11.4  \\
			&   100 &   9.68  & 6.8  & 10.9  \\
\end{tabular}
\end{ruledtabular}
\footnotetext{Frequencies in GHz. Beam size $\theta_{{\rm beam}}$ is the FWHM in
arcminutes. Sensitivities $\sigma_T$ and $\sigma_P$ are in $\mu K$ per FWHM beam.}
\end{table}
After the above operations, we build two mock data sets ($C_{\ell}^{TT}$, $C_{\ell}^{TE}$, $C_{\ell}^{EE}$, $C_{\ell}^{dd}$ and $C_{\ell}^{Td}$), that we shall call Mock-A and Mock-B, whose fiducial cosmological parameter values (see Table~\ref{tab:fiducialcp}) are, respectively, the best-fit values of base-$\Lambda$CDM+$A_{\rm L}$ to {\it Planck}-2015 TT+lowP and {\it Planck}-2015 TT+lowP+lensing data sets\footnote{ The best-fit values are read from the PLA-PR2-2015 official chains \texttt{base-Alens-plikHM-TT-lowTEB} and \texttt{base-Alens-plikHM-TT-lowTEB-lensing} at \url{http://pla.esac.esa.int/pla/}.}. 
Since the Mock-A data set, which mimics {\it Planck}-2015 TT+lowP, is generated from $A_{\rm L}\sim1.3$, we can treat it as a realization of a non-$\Lambda$CDM universe; Mock-B data, that mimics {\it Planck}-2015 TT+lowP+lensing, $A_{\rm L}\sim1.0$, is closer to a realization of $\Lambda$CDM universe. 

Based on Mock-A and Mock-B data sets, we build a ``tension" data set, called Mock-C, by combining ($C_{\ell}^{TT}$, $C_{\ell}^{TE}$, $C_{\ell}^{EE}$) from Mock-A and ($C_{\ell}^{dd}$ and $C_{\ell}^{Td}$) from Mock-B. The resulting data set should mimic the data compilation of TT+lowP+lensing data sets of {\it Planck}-2015 data release. 
In the rest of the paper, we will study several modified gravity models against Mock-A and Mock-C data sets to see whether the modified gravity models can or cannot reconcile the tension between CMB anisotropy data and CMB lensing data. 

\begin{table}
\caption{\label{tab:fiducialcp}Fiducial parameters of the mock data sets}
\begin{ruledtabular}
\begin{tabular}{cccccccc}
CP 					&	Mock-A				& 	Mock-B \\
\hline
$10^9A_s$
					&	$2.10745$			&	$2.14338$ \\	
$n_s$
					&	$0.97468$			&	$0.97156$  \\
$\tau$
					&	$0.0611$				&	$0.0664$	\\
$\Omega_bh^2$
					&	$0.022674$			&	$0.022379$	\\
$\Omega_ch^2$
					&	$0.11639$				&	$0.11748$ \\
$H_0$
					&	$69.02$				&	$68.39$		\\
$A_{\rm L}$			&      $1.28$				&      $1.02$ \\
$\sum m_{\nu}/{\rm eV}$ 	&      $0.06$				&      $0.06$ \\
\end{tabular}
\end{ruledtabular}
\footnotetext[1]{Fiducial parameter values in Mock A data sets are the best-fit values of base-$\Lambda$CDM+$A_{\rm L}$ to {\it Planck}-2015 TT+lowP data sets.}\\
\footnotetext[2]{Fiducial parameter values in Mock B data sets are the best-fit values of base-$\Lambda$CDM+$A_{\rm L}$ to {\it Planck}-2015 TT+lowP+lensing data sets.}
\end{table}

\begin{figure*}[htb!]
\centering
\includegraphics[width=1.0\textwidth]{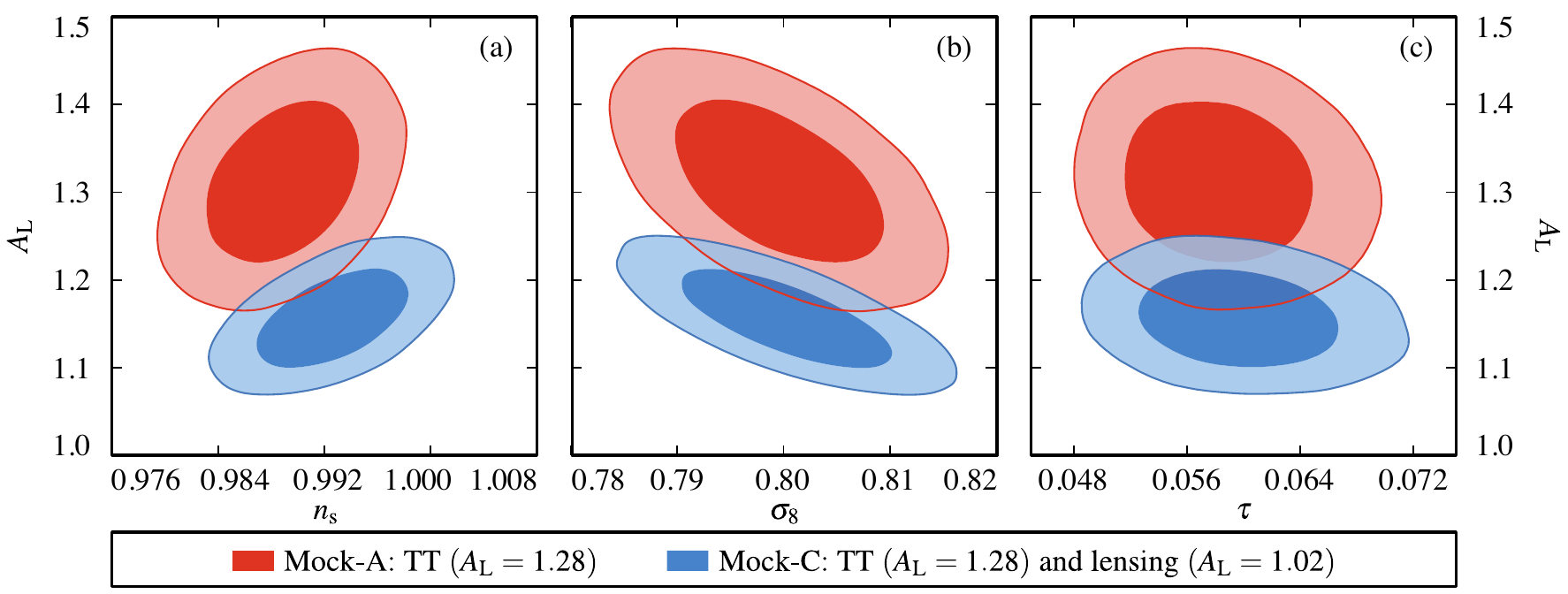}
\caption{the marginalized joint likelihood for the lensing amplitude $A_{\rm L}$ and the scalar spectral index $n_s$ {\it panel (a)}, the amplitude of the (linear) power spectrum on the scale of $8h^{-1}$Mpc, $\sigma_8$ {\it panel (b)} and the reionization optical depth $\tau$ {\it panel (c)}. 
In all three panels different colors correspond to different combination of mock data sets as shown in legend. \\
The darker and lighter shades correspond respectively to the $68\%$ C.L. and the $95\%$ C.L..}
\label{fig:AL}
\end{figure*}

\section{Modified gravity in the effective field theory framework}
The clustering of matter curves the trajectories of CMB photons and mixes the CMB photon anisotropies from different directions. Since the characteristic scale of clustering of galaxies is smaller than $10{\rm Mpc}h^{-1}$ we can use the flat sky approximation to compute the lensing contributions to the CMB anisotropy spectrum. 
The lensing signal, in general, will be encoded in all the types of CMB anisotropy modes (T,E,B), and here we only show the effect on the TT spectrum as an example, 
\begin{eqnarray}
\label{lensedTT}
C_{\ell}^{\tilde T\tilde T} &\simeq& C_{\ell}^{TT}+\int\frac{d^2\Bell'}{(2\pi)^2}\left[\Bell'\cdot(\Bell-\Bell')\right]^2C^{\phi\phi}_{|\Bell-\Bell'|}C_{\ell}^{TT}\nonumber\\
&-&C_{\ell}^{TT}\int\frac{d^2\Bell'}{(2\pi)^2}(\Bell\cdot\Bell')^2C^{\phi\phi}_{\ell'}\;,
\end{eqnarray}
in the numerical analysis that follows we take lensing effect into account also for TE and EE spectra. For the expressions of lensed TE and EE spectra we refer the reader to~\cite{Lewis:2006fu}. 

In the one extra parameter extension of the base-$\Lambda$CDM model in the {\it Planck}-2013~\cite{Ade:2013zuv} and {\it Planck}-2015~\cite{Planck:2015xua} results, the collaboration studied the case of varying the lensing amplitude parameter, $A_{\rm L}$, in the CMB anisotropies, which was originally introduced in~\cite{Calabrese:2008rt}.
This phenomenological parameter is defined by $C_{\ell}^{\phi\phi}\rightarrow A_{\rm L}C_{\ell}^{\phi\phi}$, which simply rescales the lensing amplitude contribution to the CMB anisotropies. 
This parameter, however, only modulates the CMB anisotropy spectra, $C_{\ell}^{TT}$, $C_{\ell}^{TE}$ and $C_{\ell}^{EE}$, and rescales the lensing potential spectrum $C^{\phi\phi}_{\ell}$ but does not rescale the estimator of the lensing spectrum $\hat A_{\rm L}^{\phi\phi}$ which is computed from the CMB anisotropy trispectra~\cite{Ade:2013tyw,Ade:2015zua}. 

\begin{figure*}[htb!]
\centering
\includegraphics[width=0.99\textwidth]{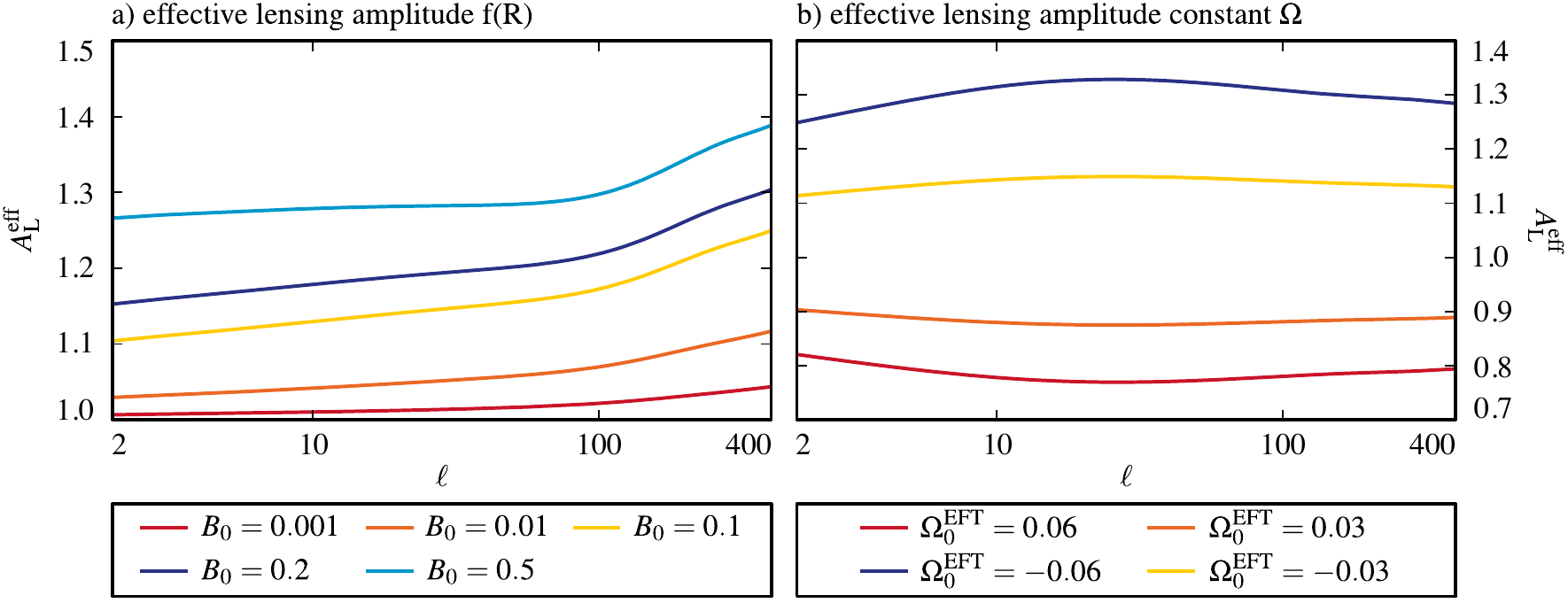}
\caption{the effective lensing amplitude $A_{\rm L}^{\rm eff}(\ell) \equiv C^{\phi\phi}_{\ell}({\rm MG})/C^{\phi\phi}_{\ell}({\rm GR})$ as a function of scale in the  modified gravity models considered in this paper. Different colors correspond to different values of the model parameters as shown in legend.}
\label{fig:ALeff}
\end{figure*}

In the following we shall elaborate on the relationship between this phenomenological parameter and modifications of gravity.
We shall consider two models that are enclosed in the background part of the EFT formalism for cosmic acceleration~\cite{Gubitosi:2012hu, Bloomfield:2012ff}.
Both can be derived from this action written in the unitary gauge and Jordan frame
\begin{align}\label{EFT_action}
  S = \int d^4x \sqrt{-g} &  \left \{ \frac{m_0^2}{2} \left[1+\Omega(\tau)\right]R+ \Lambda(\tau) \right.\nonumber\\
  &\left.- a^2c(\tau) \delta g^{00} + \ldots  \right\} + S_{m} [g_{\mu \nu}] \,,
\end{align}
where we have used conformal time and $\Omega$, $\Lambda$ and $c$ are free functions of time which multiply all the operators that are consistent with time-dependent spatial diffeomorphism invariance and contribute to the background evolution. The ellipsis indicate operators which would affect only linear and non-linear cosmological perturbations, while $S_{m}$ indicates the action for all matter fields: cold dark matter, baryons, massive and massless neutrinos and photons. 

The first modified gravity model that we use is $f(R)$, its mapping to the EFT framework was presented in~\cite{Gubitosi:2012hu} and we refer the reader to~\cite{Song:2006ej, Pogosian:2007sw, Bean:2006up, DeFelice:2010aj} for detailed discussions on the cosmology of these models.

The second model instead consist in taking a constant value for the conformal coupling $\Omega(a)=\Omega_0^{\rm EFT}$ and requiring the expansion history to be exactly that of the $\Lambda$CDM model. This requirement will then fix, through the Friedmann equations, the time dependence of the operators $c$ and $\Lambda$. 

We highlight here that the constant $\Omega$ model is not a simple redefinition of the gravitational constant.
In fact the requirement of having a $\Lambda$CDM background with a non-vanishing $\Omega$, that would change the expansion history, means that a scalar field is sourced in order to compensate this change.
This scalar field will then interact with the other matter fields and modify the behaviour of cosmological perturbations and consequently the CMB power spectra and the growth of structure. For instance, it is easy to show that in the constant $\Omega$ model, $c(\tau)$, which is vanishing in general relativity, is non-zero and reads
\begin{equation}
\label{consto_c}
c=\frac{\Omega}{2}(\rho_m+P_m)\;. 
\end{equation}
%

\begin{figure*}[htb!]
\centering
\includegraphics[width=1\textwidth]{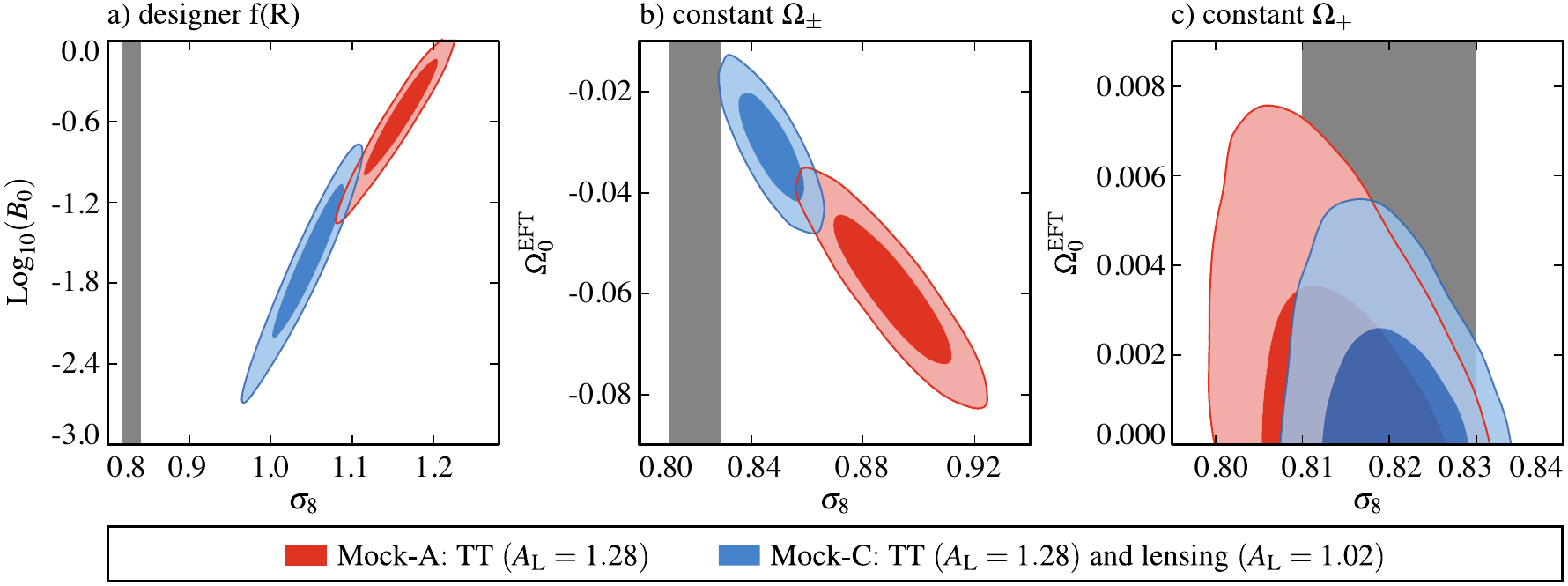}
\caption{the marginalized joint likelihood for the amplitude of the (linear) power spectrum on the scale of $8h^{-1}$Mpc, $\sigma_8$ and the present day value of ${\rm Log}_{10}B_0$ {\it panel (a)}, the present day value of the conformal coupling $\Omega_0^{\rm EFT}$ in the case in which it is allowed to have positive and negative values {\it panel (b)} and in the case in which it is restricted to positive values {\it panel (c)}.
In all three panels different colors correspond to different combination of mock data sets as shown in legend and the grey band is the marginalized $1\sigma$ bound on $\sigma_8$ from the base-$\Lambda$CDM+$A_{\rm L}$ model.
The darker and lighter shades correspond respectively to the $68\%$ C.L. and the $95\%$ C.L..}
\label{fig:mg}
\end{figure*}

Another general remark we would like to make on the models that we consider here, is that they display a radically different cosmology, as they correspond to two different behaviour of the perturbation's effective gravitational constant. 
Viable models, in the $f(R)$ case~\cite{Pogosian:2007sw}, correspond to an enhancement of the gravitational constant which in turn results in the amplification of the growth of structure that enhances substantially the lensing of the CMB. \\
In the second case we consider two possibilities. If the constant $\Omega$ is positive the model will be characterized by a smaller effective gravitational constant resulting in a suppression of the growth and consequently a suppression of the CMB lensing. We shall call this case the $\Omega_+$ model. 
If the constant $\Omega$ is negative, on the other hand, the model will have an enhanced effective gravitational constant with a phenomenology similar to that of $f(R)$ models. In contrast to what happens to the $\Omega_+$ case, that respects all the usual requirements of physical viability~\cite{Hu:2013twa}, this model is only classically stable. This means that perturbations around the FRW background are stable and well behaved but, for example, the sign of the scalar field kinetic term is wrong. 
We shall call the case in which the constant $\Omega$ can be greater and smaller than zero the $\Omega_{\pm}$ model.

To study the phenomenology of these three models we use the EFTCAMB code \footnote{Publicly available at: \url{http://wwwhome.lorentz.leidenuniv.nl/~hu/codes/}.}~\cite{Hu:2013twa, Raveri:2014cka} that allows to compute cosmological observables for all the theories enclosed in the EFT formalism once a precise mapping or parametrization is given.
An important feature of EFTCAMB, besides its versatility, is that it evolves the full perturbation equations, on all linear scales, without relying on any quasi-static approximation. In addition it checks the stability conditions of perturbations in the dark sector in order to ensure that the underlying gravitational theory is acceptable. 
The implementation details of the code and the equations that are solved can be seen in~\cite{Hu:2014oga} with further comments on the models that we are considering. 

\section{Results and discussions}
We firstly check the reliability of our simulations. As reported in the Table~\ref{tab:fiducialcp} we use the fiducial $A_{\rm L}=1.28$ for the generation of the Mock-A data set. After going through the Markov-Chain Monte-Carlo analysis~\cite{Lewis:2002ah}, we get the marginalized constraint from the Mock-A data set as
\begin{equation}
A_{\rm L}=1.31\pm0.06\;, (68\%{\rm C.L.})\;.
\end{equation}
The consistency between the input and output values of our simulations can also be seen in the panels of Figure~\ref{fig:AL}.
In addition from panel (a) we can see that there is a relatively large positive correlation between $n_s$ and $A_{\rm L}$. This happens because a large value of $n_s$ will enhance the high-$\ell$ multipoles, while an increase in $A_{\rm L}$ will smear the peaks in the same multipole range.
Panel (b) instead shows that $\sigma_8$ and $A_{\rm L}$ are anti-correlated. An increase of $A_{\rm L}$ will suppress the growth inferred from the TT power spectrum, hence we end up with lower $\sigma_8$ value.
Panel (c), at last, shows that no significant degeneracy arises between $A_{\rm L}$ and the re-ionization optical depth parameter $\tau$. 

From all the panels of Figure~\ref{fig:AL} we can see how different cosmological parameters reacts to the use of a tension data set.  
The shift of blue and red contour from Mock-A and Mock-C data sets shows the tension between them. In particular in panel (a) we can see that the Mock-C data set gives a value of $A_{\rm L}$ that is in the middle of the Mock-A and Mock-B data sets that were used to build it. From panel (b) and (c) we can instead see that the marginal probability distribution of $\sigma_8$ and $\tau$ does not change significantly as the tension in the data set is introduced. This happens because the constraint on these two quantities are dominated by the TT and EE power spectra.

\begin{table}
\caption{\label{tab:chisq} The best-fit $\chi^2$}
\begin{ruledtabular}
\begin{tabular}{cccccccc}
$\chi^2_{\rm best-fit}$ & Mock-A & Mock-C \\
\hline
base-$\Lambda$CDM		& 1967.373   & 3314.547 \\
base-$\Lambda$CDM+$A_{\rm L}$	& 1951.122   & 3303.599  \\
$f(R)$				& 1952.325   & 3305.251  \\
$\Omega_+$			& 1967.604   & 3314.676  \\
$\Omega_{\pm}$			& 1951.109   & 3304.507  \\
\end{tabular}
\end{ruledtabular}
\end{table}
After checking the consistency of our simulations we move to their interpretation within the modified gravity models here considered.

First of all we check whether these models can reproduce the input amplitude of lensing. To do so we define an effective lensing amplitude as $A_{\rm L}^{\rm eff}(\ell)\equiv C^{\phi\phi}_{\ell}({\rm MG})/C^{\phi\phi}_{\ell}({\rm GR})$ and we plot it for several choices of the parameters defining our modified gravity models in Figure~\ref{fig:ALeff}. 
From both panels we can see that the effective lensing amplitude is generally a function of scale so that the accuracy of using a scale independent approximation for it is limited by the accuracy of observations. That is, if observations are not precise enough then the scale independent approximation can work while if the observations have enough sensitivity we would bias parameter estimation by considering it scale independent. 
This point will be more and more relevant as the experimental accuracy improves. 

The $f(R)$ case, shown in panel (a), in particular, displays an $A^{\rm eff}_{\rm L}$ that is monotonically increasing as a function of scale. At low multipoles the agreement with a scale independent model is as good as few percent while in the high-$\ell$ range it is off by $40\%$ depending on the present value of the Compton wavelength of the scalaron. \\
The constant $\Omega_{+}$ and $\Omega_{\pm}$ models are shown in panel (b). Unlike the $f(R)$ case, $A^{\rm eff}_{\rm L}(\ell)$ has a wide bump or dip around multipoles of few tenths while the scale dependence is somewhat weaker than the previous case.
As expected, in the $\Omega_+$ model the amplitude of lensing is decreased as a consequence of the suppression of growth while the opposite behaviour shows when $\Omega_0^{\rm EFT}$ is smaller than zero. 

We fit these models to our Mock-A and Mock-C data sets and we show the marginalized bound on parameters of interest in Figure~\ref{fig:mg}. 
From panel (a) we can see that there is a strong positive correlation between the scalaron present Compton wavelength parameter $B_0$ and $\sigma_8$. This effect is well known and corresponds to the fact that as the effective Newton constant is increased the growth of matter perturbations is increased as well. 
When considering just Mock-A our results show that in order to mimic $A_{\rm L}\sim 1.3$, the marginalized constraint on $B_0$ have to point significantly toward large values of the scalaron Compton wavelength. The marginal bound is then $-1.15 < {\rm Log}_{10}B_0 <  -0.04$ at $95\%$ C.L.. 

Due to the significant degeneracy between $B_0$ and $\sigma_8$ such values of $B_0$ will lead to strong enhancement of the growth rate at relatively small scales making the $\sigma_8$ value too large so that the tension between {\it Planck} and and LSS surveys, such as CFHTLenS~\cite{Heymans:2012gg,Erben:2012zw}, would become even worse.
For this reason when lensing is added in Mock-C, the tension in $A_{\rm L}$, that for the lensing data set is smaller, pushes the posterior of $B_0$ toward smaller values, making it closer to GR. 
A similar effect was also observed for the {\it Planck}-2015 data set~\cite{Ade:2015rim} and in the {\it Planck}-2013 data~\cite{Hu:2014sea,Dossett:2014oia,Hu:2013aqa,Marchini:2013oya}. 

As expected a similar result is found also in the $\Omega_{\pm}$ model, as shown in panel (b). 
Here the correlation between $\Omega_0^{\rm EFT}$ is negative due to the fact that negative values of $\Omega$ correspond to significant deviations from GR, and consequently to stronger enhancement of CMB lensing.
All the conclusions, previously discussed in the $f(R)$ case, apply also to this model with the relevant exception that the enhancement of the growth is not so dramatic as in $f(R)$. The bound on $\sigma_8$ from the $A_{\rm L}$ fit, shown in Figure~\ref{fig:mg} as a gray band, is in fact almost compatible with the one in the $\Omega_{\pm}$ case.
This is due to the fact that deviations from scale independence of $A_{\rm L}^{\rm eff}$, in this model, are weaker and a constant $\Omega$ is more efficient at mimicking a scale independent $A_{\rm L}$ than $f(R)$ models.
Since a weaker effective Newton constant is disfavoured by the {\it Planck}-2015 CMB anisotropy data, the best-fit parameter in the $\Omega_+$ model mimic those in the base-$\Lambda$CDM model and the model result to be compatible with GR. This conclusion is similar to the one found in~\cite{Raveri:2014cka,Hu:2014sea,Ade:2015rim}.

At last in Table~\ref{tab:chisq} we list the best-fit $\chi^2$ for the one parameter extensions of the base-$\Lambda$CDM model that we investigated in this paper.
From there we can see that the best-fit $\chi^2$ values in the base-$\Lambda$CDM+$A_{\rm L}$, $f(R)$, $\Omega_{\pm}$ models are similar and are noticeably smaller 
than those from base-$\Lambda$CDM or $\Omega_+$ models with $\Delta\chi^2\simeq -16$ from the Mock-A and $\Delta\chi^2\simeq -11$ from the Mock-C data sets.
The best-fit $\chi^2$ in the $\Omega_{\pm}$ case is slightly better than the $f(R)$ case and closer to the $A_{\rm L}$ one because of the weaker scale dependence of $A^{\rm eff}_{\rm L}$. \\

In conclusion, {\it Planck}-2015 results revealed some interesting tensions between CMB temperature and polarization anisotropies and CMB lensing. These tensions add to the one reported by LSS surveys, like CFHTLenS, that seem to favour a smaller $\sigma_8$. 
If this is not due to some unaccounted systematic effects, this might be an indication of exotic physics beyond the base-$\Lambda$CDM model and as such should be investigated in great detail. \\
In this paper, using a simulation of the {\it Planck}-2015 data set, we tried to reconcile this tension with some modified gravity models.
In particular we studied whether this tension can be mitigated by $f(R)$ models or models characterized by a constant conformal coupling between gravitational and matter  perturbations.
We found that the $f(R)$ and the $\Omega_{\pm}$ models can mimic the role of $A_{\rm L}$ even if, generally, the amplitude of lensing, with respect to the GR case, is scale dependent.
In order to provide a good fit to the {\it Planck}-like CMB anisotropy spectra, however, these models predict an enhancement of the growth on smaller scales that make the tension in $\sigma_8$ even worse.
From these results we can conclude that the tension between the growth of matter perturbations assessed from CMB power spectra, CMB lensing and LSS surveys can be mimicked by modified gravity models but is hardly relieved by simple models beyond the standard $\Lambda$CDM one.
This in turn suggests to investigate more complicated models, possibly with different time dependencies, to allow for different regimes of growth at the times at which each of these observations is more sensitive.

\begin{acknowledgments}
We thank Carlo Baccigalupi, Julien Lesgourgues, Matteo Martinelli, Sabino Matarrese, Laurence Perotto, Alessandra Silvestri, Wessel Valkenburg and Matteo Viel for helpful correspondences and discussions. 
B.H. is supported by the Dutch Foundation for Fundamental Research on Matter (FOM). M.R. acknowledges partial support from the INFN-INDARK initiative.
M.R. thanks the Instituut Lorentz (Leiden University) for hospitality while this work was being completed.
\end{acknowledgments}


\begin{thebibliography}{999}

\bibitem{Planck:2015xua} 
  P.~A.~R.~Ade {\it et al.}  [Planck Collaboration],
  arXiv:1502.01589 [astro-ph.CO].
  
\bibitem{Perotto:2006rj} 
  L.~Perotto, J.~Lesgourgues, S.~Hannestad, H.~Tu and Y.~Y.~Y.~Wong,
  JCAP {\bf 0610}, 013 (2006)
  [astro-ph/0606227].
  
\bibitem{Planck:2006aa} 
  J.~Tauber {\it et al.}  [Planck Collaboration],
  astro-ph/0604069.
  
\bibitem{Hu:2001kj} 
  W.~Hu and T.~Okamoto,
  Astrophys.\ J.\  {\bf 574}, 566 (2002)
  [astro-ph/0111606].
  
\bibitem{Lewis:1999bs} 
  A.~Lewis, A.~Challinor and A.~Lasenby,
  Astrophys.\ J.\  {\bf 538}, 473 (2000),
  [astro-ph/9911177].
  
\bibitem{CAMB}
http://camb.info \,.  

\bibitem{Tegmark:1996bz} 
  M.~Tegmark, A.~Taylor and A.~Heavens,
  Astrophys.\ J.\  {\bf 480}, 22 (1997)
  [astro-ph/9603021].
  
\bibitem{Lewis:2005tp} 
  A.~Lewis,
  Phys.\ Rev.\ D {\bf 71}, 083008 (2005)
  [astro-ph/0502469].
  
\bibitem{Lewis:2006fu} 
  A.~Lewis and A.~Challinor,
  Phys.\ Rept.\  {\bf 429}, 1 (2006)
  [astro-ph/0601594].
  
\bibitem{Ade:2013zuv} 
  P.~A.~R.~Ade {\it et al.}  [Planck Collaboration],
  Astron.\ Astrophys.\  {\bf 571}, A16 (2014)
  [arXiv:1303.5076 [astro-ph.CO]].
  
\bibitem{Calabrese:2008rt} 
  E.~Calabrese, A.~Slosar, A.~Melchiorri, G.~F.~Smoot and O.~Zahn,
  Phys.\ Rev.\ D {\bf 77}, 123531 (2008)
  [arXiv:0803.2309 [astro-ph]].

\bibitem{Ade:2015zua} 
  P.~A.~R.~Ade {\it et al.}  [Planck Collaboration],
  arXiv:1502.01591 [astro-ph.CO].
  
\bibitem{Ade:2013tyw} 
  P.~A.~R.~Ade {\it et al.}  [Planck Collaboration],
  Astron.\ Astrophys.\  {\bf 571}, A17 (2014)
  [arXiv:1303.5077 [astro-ph.CO]].
  
\bibitem{Gubitosi:2012hu} 
  G.~Gubitosi, F.~Piazza and F.~Vernizzi,
  JCAP {\bf 1302}, 032 (2013)
  [JCAP {\bf 1302}, 032 (2013)]
  [arXiv:1210.0201 [hep-th]].
  
\bibitem{Bloomfield:2012ff} 
  J.~K.~Bloomfield, \'E. \'E. ~Flanagan, M.~Park and S.~Watson,
JCAP {\bf 1308}, 010  (2013),
  [arXiv:1211.7054 [astro-ph.CO]].    
  
\bibitem{Song:2006ej} 
  Y.~S.~Song, W.~Hu and I.~Sawicki,
  Phys.\ Rev.\ D {\bf 75}, 044004 (2007)
  [astro-ph/0610532].
  
\bibitem{Pogosian:2007sw} 
  L.~Pogosian and A.~Silvestri,
  Phys.\ Rev.\ D {\bf 77}, 023503 (2008)
  [Erratum-ibid.\ D {\bf 81}, 049901 (2010)]
  [arXiv:0709.0296 [astro-ph]].
  
\bibitem{Bean:2006up} 
  R.~Bean, D.~Bernat, L.~Pogosian, A.~Silvestri and M.~Trodden,
  Phys.\ Rev.\ D {\bf 75}, 064020 (2007)
  [astro-ph/0611321].
  
\bibitem{DeFelice:2010aj} 
  A.~De Felice and S.~Tsujikawa,
  Living Rev.\ Rel.\  {\bf 13}, 3 (2010)
  [arXiv:1002.4928 [gr-qc]].
  
\bibitem{Hu:2013twa} 
  B.~Hu, M.~Raveri, N.~Frusciante and A.~Silvestri,
  Phys.\ Rev.\ D {\bf 89}, no. 10, 103530 (2014)
  [arXiv:1312.5742 [astro-ph.CO]].

\bibitem{Raveri:2014cka} 
  M.~Raveri, B.~Hu, N.~Frusciante and A.~Silvestri,
  Phys.\ Rev.\ D {\bf 90}, no. 4, 043513 (2014)
  [arXiv:1405.1022 [astro-ph.CO]].
  
\bibitem{Hu:2014oga} 
  B.~Hu, M.~Raveri, N.~Frusciante and A.~Silvestri,
  arXiv:1405.3590 [astro-ph.IM].
 
\bibitem{Lewis:2002ah} 
  A.~Lewis and S.~Bridle,
  Phys.\ Rev.\ D {\bf 66}, 103511 (2002)
  [astro-ph/0205436].
 
\bibitem{Heymans:2012gg} 
  C.~Heymans, L.~Van Waerbeke, L.~Miller, T.~Erben, H.~Hildebrandt, H.~Hoekstra, T.~D.~Kitching and Y.~Mellier {\it et al.},
  Mon.\ Not.\ Roy.\ Astron.\ Soc.\  {\bf 427}, 146 (2012)
  [arXiv:1210.0032 [astro-ph.CO]].
  
\bibitem{Erben:2012zw} 
  T.~Erben, H.~Hildebrandt, L.~Miller, L.~van Waerbeke, C.~Heymans, H.~Hoekstra, T.~D.~Kitching and Y.~Mellier {\it et al.},
  Mon.\ Not.\ Roy.\ Astron.\ Soc.\  {\bf 433}, 2545 (2013)
  [arXiv:1210.8156 [astro-ph.CO]].
   
\bibitem{Ade:2015rim} 
  P.~A.~R.~Ade {\it et al.}  [XXX Collaboration],
  arXiv:1502.01590 [astro-ph.CO].
 
\bibitem{Hu:2014sea} 
  B.~Hu, M.~Raveri, A.~Silvestri and N.~Frusciante,
  arXiv:1410.5807 [astro-ph.CO].
  
\bibitem{Dossett:2014oia} 
  J.~Dossett, B.~Hu and D.~Parkinson,
  JCAP {\bf 1403}, 046 (2014)
  [arXiv:1401.3980 [astro-ph.CO]].
  
\bibitem{Hu:2013aqa} 
  B.~Hu, M.~Liguori, N.~Bartolo and S.~Matarrese,
  Phys.\ Rev.\ D {\bf 88}, no. 12, 123514 (2013)
  [arXiv:1307.5276 [astro-ph.CO]].
  
\bibitem{Marchini:2013oya} 
  A.~Marchini and V.~Salvatelli,
  Phys.\ Rev.\ D {\bf 88}, no. 2, 027502 (2013)
  [arXiv:1307.2002 [astro-ph.CO]].
  
\end{thebibliography}
\end{document}